\documentclass[letterpaper,10pt,twocolumn]{article}   
\usepackage{ol2}
\usepackage[draft]{hyperref} 
\def\lsim{\mathrel{\rlap{\lower4pt\hbox{\hskip1pt$\sim$}}
    \raise1pt\hbox{$<$}}}                
\def\gsim{\mathrel{\rlap{\lower4pt\hbox{\hskip1pt$\sim$}}
    \raise1pt\hbox{$>$}}}                

\usepackage{amsmath}
\usepackage{amssymb}
\usepackage{color}

\begin{document}
\def\jlt{ J.\ Lightwave\ Technol.\ }
\def\ptl{ IEEE Photon.\ Tech.\ Lett.\ }
\def\omex{ Opt.\ Mat.\ Express }
\def\pr{Phys.\ Rev. }

\def\gammat{\tilde\gamma}
\def\etat{\tilde\eta}
\def\Pt{\tilde P}
\def\phit{\tilde\phi}
\def\Xt{\tilde X}
\def\Lt{\tilde L}
\def\rt{\tilde r}
\def\LG{\rm LG}
\def \NLSE {nonlinear Schr\"{o}dinger equation}

\newcommand{\ket}[1]{\left| #1 \right>} 
\newcommand{\bra}[1]{\left< #1 \right|} 
\newcommand{\braket}[2]{\left< #1 \vphantom{#2} \right|
 \left. #2 \vphantom{#1} \right>} 

\twocolumn[ 

\title{Nonlinear multimodal interference and saturable absorption using a short graded-index multimode optical fiber}


\author{Elham Nazemosadat,$^{1}$ and Arash Mafi$^{1,*}$}
\address{$^1$Department of Electrical Engineering and Computer Science, \\ University of Wisconsin-Milwaukee, Milwaukee, WI 53211, USA}
\address{$^*$Corresponding author: mafi@uwm.edu}

\begin{abstract}
{ A detailed investigation of the nonlinear multimodal interference in a short graded-index multimode optical fiber is presented.}
The analysis is performed for a specific device geometry, where the light is coupled in and out of the multimode fiber
via single-mode fibers. The same device geometry was recently used to obtain ultra-low-loss coupling between two 
single-mode optical fibers with very different mode-field diameters. Our results indicate the potential application of this simple geometry for
nonlinear devices, such as in nonlinear switching, optical signal processing, or as saturable absorbers in mode-locked 
fiber lasers. { Saturable absorption in this all-fiber configuration is discussed and it is shown that
it provides attractive properties that can potentially be used in high pulse energy mode-locked fiber lasers.}   
\end{abstract}


] 

\maketitle 

\section{Introduction}
\begin{figure}[t]
\centering
\includegraphics[width=2.5in]{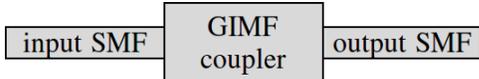}
\caption{A GIMF of length $L$ is used as an intermediate coupler between two SMF fibers.
In Refs.~\cite{MafiMMI1,MafiMMI2}, this geometry was used to create very low-loss couplers 
between two SMFs with very different mode-field diameters. In this paper, 
nonlinear MMI effects for identical input and output SMFs { are explored}.}
\label{fig:coupler}
\end{figure}

Multimode interference (MMI) in optical fibers has been used successfully 
in recent years for various device applications including, beam shapers, 
sensors, and filters~\cite{Mehta,Mohammed1,Mohammed2,Wang,Yilmaz,Zhu0,Zhu1,Zhu2}. 
{ MMI in a graded-index multimode optical fiber (GIMF) was recently used to}
create very low-loss couplers between two single mode optical fibers (SMFs) with very 
different mode-field diameters~\cite{MafiMMI1,MafiMMI2}. 
{ Here, the linear analysis of Ref.~\cite{MafiMMI1} is extended and the MMI 
effect in the nonlinear regime for the GIMF-coupler geometry shown in Fig.~\ref{fig:coupler} is investigated.} 
The main intention of this study is to explore the possibility of using the 
simple SMF-GIMF-SMF geometry in nonlinear device applications, such as in optical 
signal processing~\cite{AgrawalBook} or as a saturable absorber (SA) in mode-locked fiber 
lasers~\cite{1stmodelocked,2ndmodelocked}.

GIMFs are commonly used in fiber-optic communications to reduce modal dispersion, as
the group velocities of all modes are nearly identical at the design wavelength~\cite{Gloge,Okamoto}.
However, GIMFs exhibit another unique property that makes them very attractive for MMI applications:
the propagation constants of their modes are equally spaced. 
Consequently, their self-imaging lengths can be very short, even less than 1~mm~\cite{MafiMMI1}; therefore, 
it is possible to make extremely short (even submillimeter) practical MMI devices~\cite{MafiMMI2} or easily 
tune the length of the GIMF coupler for a specific MMI application~\cite{MafiMMI2}.

In this paper, { our studies are focused}
on the simplest case of nonlinear MMI (NL-MMI) in the 
setup shown in Fig.~\ref{fig:coupler} with identical input and output SMFs; this choice is
also the most practical one for most device applications. Our results will elucidate some 
general behaviors of such couplers in the nonlinear regime. More complex geometries
involving different fiber junctions and using various mode conversion techniques~\cite{Farahi} such as 
long-period gratings~\cite{Ramachandran1} can modify some of the observations and conclusions;
however, such complex modifications are less likely to be adopted in practical device applications in the near future.
We hope that our results can be a useful starting point for future studies of NL-MMI and mode conversion 
in more complex systems. 

NL-MMI has been studied in various contexts over the years (see, for example, Ref.~\cite{NLMMI1}). Generally 
speaking, nonlinear devices that operate based on nonlinear mode switching and coupling in spatially separated waveguides 
can also be viewed as MMI devices~\cite{Stegeman,Winful,Proctor,Christodoulides}. Nonlinear polarization rotation
is another important example of NL-MMI (between two orthogonal polarization modes) that closely resembles 
our analysis here~\cite{Stolen,AgrawalNLBook}.

In Section~\ref{sec:GeneralOverview}, 
a general overview of GIMFs, nonlinear propagation in multimode optical fibers, and MMI { are presented}. NL-MMI in GIMFs in the context of the SMF-GIMF-SMF geometry of 
Fig.~\ref{fig:coupler} is discussed in Section~\ref{sec:NLMMI}. In Subsection~\ref{sec:2modes},
{ a reduced version of the model which includes only two propagating modes is analyzed;}
this simplification helps us to develop proper insight into the physics of NL-MMI in GIMFs.
In Subsection~\ref{sec:5modes},
our analysis { is extended to the realistic case of five propagating modes; the impact of the 
presence of additional propagating modes in the NL-MMI setup can be shown by comparing these results with the results of Subsection~\ref{sec:2modes}.} 
In Section~\ref{sec:saturable},
the saturable absorption behavior of the SMF-GIMF-SMF geometry { is analyzed} as an example of the utility of this setup
for device applications. 
The concluding remarks { are presented} in Section~\ref{sec:conclusion},
{ where it is shown} that the SMF-GIMF-SMF geometry can be a viable design for nonlinear switching or saturable absorption.

The formalism and results in this paper are mainly laid out in dimensionless units; this choice is common 
in nonlinear fiber optics, because it reduces the number of parameters and prevents redundancy in the analysis.
For real-world applications, it is easy to convert back to the dimensionful parameters; in Section~\ref{sec:conclusion},
the performance of a NL-MMI SA device with specific parameters from commercially available optical fibers { is analyzed}. 

{ In order to reduce the complexity of the analysis and to make the problem more tractable, 
the studies in this paper are considered in the continuous wave (CW) limit.} 
In practice, after the design parameters are chosen based on the CW analysis, the temporal effects can be included to optimize the 
final design. GIMFs, if used near their optimal design wavelengths, are particularly attractive 
for short-pulse ultrafast applications compared with other MMFs. The low modal dispersion in GIMFs ensures 
that pulses do not break up in short GIMF segments~\cite{MafiNL}, making the CW analysis an adequate approximation.

Finally, 
{ it should be pointed out} that the transmission through the SMF-GIMF-SMF geometry is a periodic function of the 
frequency of the light source~\cite{MafiMMI1,MafiMMI2}. For mode-locking applications where short pulses are 
generated with very large spectral bandwidth, the length of the GIMF needs to be sufficiently short to provide 
the necessary spectral transmission window. Moreover, the SMF-GIMF-SMF geometry can be potentially used as 
the spectral filter that is required to stabilize the mode-locking operation
of normal-dispersion high-energy femtosecond fiber lasers~\cite{Chong}.
\section{Fundamentals}
\label{sec:GeneralOverview}
In the following three subsections, 
a brief general overview of GIMFs (Subsection~\ref{sec:gimfReview}), nonlinear multimode propagation of light in 
GIMFs (Subsection~\ref{sec:NLGIMF}), and the general formulation of the  MMI phoenomenon in GIMFs in the context of the SMF-GIMF-SMF coupler (Subsection~\ref{sec:smfGIMFsmfbasics}) { will be presented}.
The formulation presented in this section { will be used} to analyze the nonlinear behavior of the SMF-GIMF-SMF 
geometry in detail in subsequent sections.
\subsection{Overview of GIMFs}
\label{sec:gimfReview}
The refractive index profile of a GIMF is given by
\begin{equation}
\label{eq:indexprofile}
n^2(\rho)=n_0^2\left[1-2\Delta\left(\frac{\rho}{R}\right)^\alpha\right],
\end{equation}
where $R$ is the core radius, $n_0$ is the refractive index at the center of the core, 
$\Delta$ is the index step, $\alpha\approx 2$ characterizes a 
near parabolic-index profile in the core ($\rho\le R$), and $\alpha = 0$ in the cladding ($\rho>R$). 
The transverse electric field profile of a confined mode, 
with radial $p$ ($p\ge 0$) and angular $m$ integer numbers,
can be expressed as~\cite{MafiNL}
\begin{equation}
\label{eq:Eprofile} 
E_{p,m}(\rho,\phi)=N^m_p
\frac{\rho^{|m|}}{\rho_0^{|m|+1}}
\exp(-\frac{\rho^2}{2\rho_0^2})
L^{|m|}_{p}(\frac{\rho^2}{\rho_0^2})
e^{im\phi},
\end{equation}
where $L^{|m|}_{p}$ are generalized Laguerre polynomials, and $\rho_0$ and $N^m_p$ are given by
\begin{equation}
\label{eq:rho0}
\rho_0=\frac{R^{1/2}}{(k_0n_0)^{1/2}(2\Delta)^{1/4}},\quad N^m_p=\sqrt{\frac{p!}{\pi(p+|m|)!}},
\end{equation}
where $k_0=2\pi/\lambda$.
The coefficients $N^m_p$ of these Laguerre-Gauss modes ($\LG_{pm}$) are chosen such that the modes are 
orthonormal. Using the bra-ket notation, 
the electric field profile of the $\LG_{pm}$
mode { is identified} as $\ket{p,m}=E_{p,m}(\rho,\phi)$ and 
the orthonormality condition { is expressed} as 
$\braket{p,m}{p^\prime,m^\prime}=\delta_{p,p^\prime}\delta_{m,m^\prime}$, where the bra-ket indicates integration in the 
transverse ($\rho$-$\phi$) coordinates. 

All modes with equal group mode number
$g=2p+|m|+1$ are almost degenerate in the value of the propagation constant,
which is given according to the formula
\begin{equation}
\label{eq:beta0th}
\beta_{(g)}=n_0 k \left(1-2\Delta X_g\right)^{1/2},
\end{equation}
where
\begin{equation}
\label{eq:xdef}
X_g=\left(\frac{g}{\sqrt{N_\alpha}}\right)^{2\alpha/(\alpha+2)}.
\end{equation}
$N_\alpha$ is the total number of guided modes (counting the polarization degeneracy) and is given by
\begin{equation}
\label{eq:Nalpha}
N_\alpha=\frac{\alpha}{\alpha+2}n_0^2k_0^2R^2\Delta.
\end{equation}

While the analysis presented in this article is generally applicable to a wide range of
GIMFs and is presented in dimensionless units, for specific numerical arguments and comparisons, 
the parameters of a conventional high-bandwidth commercial-grade GIMF such as Corning's InfiniCor\textsuperscript{\textregistered} eSX+ { will be used; this fiber is}
optimized for high bandwidth performance at 850~nm wavelength, with the core radius of $R=25~\mu m$.
{ The specified GIMF will be referred to as C-GIMF standing for the {\em conventional GIMF}.}
{ It should be noted that} the key spatial dimension in a GIMF that sets all its modal properties is not the core radius $R$, 
but is $\rho_c=\sqrt{2}\rho_0$, where $\rho_0$ is defined in Eq.~\ref{eq:rho0}.
$\rho_c$ is the mode radius of the $\LG_{00}$ mode; for C-GIMF, 
{ $\rho_c\approx 7.7~\mu m$ at 1550~nm}
wavelength and $\rho_c\approx 5.7~\mu m$ at 850~nm wavelength.

\subsection{Overview of nonlinear propagation in GIMFs}
\label{sec:NLGIMF}
Consider an input optical field injected into the core of the GIMF. 
The injected field excites the $\LG_{pm}$ modes of the GIMF with different 
amplitudes $A_{p,m}(0)$; $A_{p,m}(z)$ will be regarded as the envelope of the electric field and 
$z=0$ indicates the input end of the GIMF{~\cite{MafiNL}}. In order to simplify our notation, 
the collective index $\mu$ { will be used} to represent the $p,m$ index pair of the $\LG_{pm}$ modes; i.e., $\mu\equiv (p,m)$ in the rest 
of this section. However, when necessary, we will directly use the $(p,m)$ labels or specify how to translate the $\mu$
index into the $(p,m)$ index pair. 
{ $|A_\mu(z)|^2$ represents the optical power in the $\LG_{pm}$ mode.}

In the scalar approximation, the generalized nonlinear Schr\"{o}dinger equation (GNLSE) describing the 
CW longitudinal evolution of $A_{p,m}(z)$ can be written as
\begin{align}
\label{eq:masterNL}
{
\dfrac{\partial A_\mu}{\partial z}=
i\delta\beta_{\mu}A_\mu+i\gamma \sum_{\nu,\kappa,\xi}\etat_{\mu\nu\kappa\xi}A_\nu A_\kappa A^\star_\xi.
}
\end{align}
where { $\delta\beta_{\mu}=\beta_{\mu}-\beta_{(1)}$,}  $\beta_{\mu}$ is the propagation constant of
the mode with the collective index $\mu$ and $\beta_{(1)}$ is the propagation constant of the
$\LG_{00}$ mode with $g=1$ from Eq.~\ref{eq:beta0th}.
The ``normalized'' nonlinear coupling coefficient is a fully symmetric tensor and is
defined as {$\etat_{\mu\nu\kappa\xi}=\gamma_{\mu\nu\kappa\xi}/\gamma$} where 
\begin{align}
\label{eq:gammadef}
{
\gamma_{\mu\nu\kappa\xi}=\left(\dfrac{n_2\omega_0}{c}\right)\int{d^2x} E^\star_\mu E_\nu E_\kappa E^\star_\xi.
}
\end{align}
In Eq.~\ref{eq:gammadef}, $E_\mu$ is the shorthand notation for $E_{p,m}(\rho,\phi)$. $\gamma=\gamma_{0000}$ is the 
nonlinear coefficient of the $\LG_{00}$ mode, $n_2$ is the nonlinear index coefficient, and $\omega_0$ is the carrier frequency.
For the lowest order mode ${\rm LG}_{00}$ for which $\rho_c$ is simply the modal radius, 
{ one can define} $\gamma_{0000}=n_2\omega_0/A^0_{\rm eff}$, where $A^0_{\rm eff}=\pi\rho^2_c$.
{ The total optical power} in the GIMF is given by $\Pt=\sum_{\mu}|A_{\mu}(z)|^2$
and is conserved in propagation along the GIMF; i.e., $\partial_z \Pt=0$.

It is more convenient to express the GNLSE in dimensionless units. 
$B_{\mu}$ { can be defined as}
\begin{align}
\label{eq:redef}
B_{\mu}(z)&=\dfrac{1}{\sqrt{\Pt}}A_{\mu}(z)e^{-i\gamma \Pt z},
\end{align}
and rescale the longitudinal coordinate $z$ by the difference between the propagation constants of the first and the
second mode groups $\beta_{(1)}-\beta_{(2)}$,
\begin{align}
\label{eq:zredef}
\zeta=z\times \left(\beta_{(1)}-\beta_{(2)}\right).
\end{align}
Using these transformations, Eq.~\ref{eq:masterNL} can be simplified as
\begin{align}
\label{eq:masterNL2}
{
\partial_\zeta B_\mu=-i(r_\mu+\gammat)B_\mu
+i\gammat
\sum_{\nu,\kappa,\xi}\etat_{\mu\nu\kappa\xi}B_\nu B_\kappa B^\star_\xi,
}
\end{align}
where Eq.~\ref{eq:masterNL2} is expressed in terms of the dimensionless coefficients  
\begin{align}
\label{eq:rmugammat}
r_\mu=\dfrac{\beta_{\mu}-\beta_{(1)}}{\beta_{(2)}-\beta_{(1)}},\qquad
\gammat=\dfrac{\gamma \Pt}{\beta_{(1)}-\beta_{(2)}}.
\end{align}
and can be solved for the dimensionless field $B_\mu$ in terms of the dimensionless longitudinal coordinate $\zeta$.
{ It can be realized }that the rescaled fields $B_\mu$ satisfy the power conservation condition $\sum_{\mu}|B_{\mu}(z)|^2=1$ at any points along the GIMF.

Using Eq.~\ref{eq:beta0th}, Eq.~\ref{eq:xdef}, and the fact that $X_g\ll 1$, one can show that
$r_\mu$ is non-negative and is nearly an integer: 
{ $r_\mu\approx g_\mu-1$}, where $g_\mu$ 
is the group number associated with mode $\mu$~\cite{MafiMMI2}. 
\subsection{Overview of MMI in GIMFs}
\label{sec:smfGIMFsmfbasics}

The particular setup that 
{ is considered in this paper} is shown in Fig.~\ref{fig:coupler} and
consists of injecting a nearly Gaussian beam from the input SMF, which is spliced to the input 
facet of the GIMF, and collecting the light at the other end of the GIMF from the output SMF. 
If 
the normalized mode of the input SMF { is defined as} $\ket{{\rm in}}$, where $\braket{{\rm in}}{{\rm in}}=1$, the mode amplitude in the GIMF at the input can be
written as $A_\mu(0)=\sqrt{\Pt}\braket{p,m}{{\rm in}}$, where $\ket{p,m}$ is the GIMF mode defined in Eq.~\ref{eq:Eprofile}.

The relative transmitted power to the output SMF is given by 
\begin{equation}
\tau=\dfrac{1}{\Pt}|\braket{{\rm out}}{E(\rho,\phi,L)}|^2,
\end{equation}
where $\ket{{\rm out}}$ is the normalized mode of the output SMF, and $\ket{E(\rho,\phi,L)}$ is
the total field at the output facet of the GIMF with a length of $L$, calculated from 
\begin{equation}
\label{eq:beamexpansion}
E(\rho,\phi,L)=e^{i\beta_{(1)}L}\sum_\mu{A_\mu(L) E_\mu(\rho,\phi)}.
\end{equation}
Identical input and output SMFs are the focus of this paper ($\ket{{\rm in}}\equiv \ket{{\rm out}}$), for which 
the relative power transmission can be written as
\begin{equation}
\tau=\dfrac{1}{\Pt^2}|\sum_\mu A_\mu^\star(0)A_\mu(z)|^2=|\sum_\mu B_\mu^\star(0)B_\mu(z)|^2.
\label{eq:tauSMFGIMFSMF}
\end{equation}

The injected beam from the input SMF is in the form of a Gaussian with radius $w$ and can be expressed as
\begin{equation}
\label{eq:EprofileSMin}
\ket{{\rm in}}=\sqrt{\frac{2}{\pi w^2}}\exp{\left(-\dfrac{\rho^2}{w^2}\right)}.
\end{equation}
{ Because of the azimuthal symmetry of the input Gaussian beam, if the input SMF is centrally aligned with the GIMF,  only $\LG_{p0}$ ($m=0$) modes can be excited
in the GIMF. The excitation amplitude can be calculated as~\cite{MafiMMI1}}
\begin{equation}
\label{eq:overlap}
B_{p}(0)=\dfrac{2\sqrt{\eta}}{\eta+1}\Psi^p,\quad 
\eta=\dfrac{\rho_c^2}{w^2},
\quad \Psi=\dfrac{\eta-1}{\eta+1}.
\end{equation}

\begin{figure}[htb]
\centering
\includegraphics[width=2.2in]{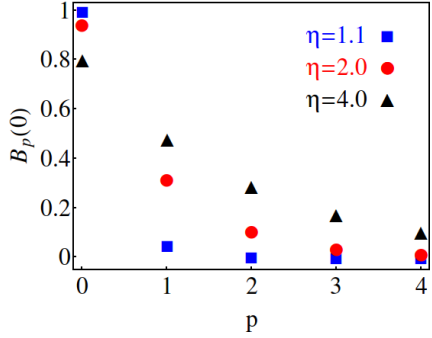}
\caption{The excitation amplitudes of the $\LG_{p0}$ modes {from Eq.~\ref{eq:overlap} are plotted} 
as a function of the radial number $p$, for three different values of the $\eta$ parameter.}
\label{fig:amplitude-vs-eta}
\end{figure}
In Fig.~\ref{fig:amplitude-vs-eta}, 
the excitation amplitudes of the  $\LG_{p0}$ modes { are displayed} 
as a function of the radial number $p$, for three different values of the $\eta$ parameter.
For the case of $\eta=1$ (not shown in Fig.~\ref{fig:amplitude-vs-eta}), only the $\LG_{00}$
mode is excited, because the input Gaussian beam is mode-matched to the $\LG_{00}$ mode of the
GIMF. For a slightly larger value of $\eta=1.1$, only 0.2\% of power is coupled to the $\LG_{01}$
mode and the power coupled to higher order modes is negligible. For the larger value of $\eta=4$,
only 64\% of the power is coupled to the $\LG_{00}$ mode. In this case, the relative power coupled 
to the $\LG_{01}$, $\LG_{02}$, and $\LG_{03}$ modes are 23\%, 8\%, and 3\%, respectively. 
{ $\eta=4$ is obtained when coupling}
 the C-GIMF to an SMF with the mode-field diameter 
of nearly $7.7~\mu m$ at 1550~nm wavelength. For coupling of the C-GIMF to Corning\textsuperscript{\textregistered}
SMF-28\textsuperscript{TM} (SMF-28) with the mode-field diameter
of $10.4~\mu m$ at 1550~nm wavelength, the value of $\eta$ is nearly equal to $2.2$.  

{ There are two reasons for not exploring the case of $\eta< 1$ in this manuscript.}
First, under the $\eta\to\eta^{-1}$ transformation, Eq.~\ref{eq:overlap}
remains unchanged except for an overall multiplicative factor of $(-1)^p$~\cite{MafiMMI1}.
Despite this change in the the initial phase, no new insight is obtained in overall observations
and general conclusions presented in this paper. Second, $\eta$ is larger than unity 
for most combinations of commercially available SMFs and GIMFs.
Therefore, 
our analysis { are limited} to $\eta>1$ to avoid adding unnecessary complexity.
\section{NL-MMI: SMF-GIMF-SMF geometry}
\label{sec:NLMMI}
In the following, 
our studies { will only include} the case of $m=0$ or zero-angular momentum modes (see the discussion in
Subsection~\ref{sec:smfGIMFsmfbasics}).
The experimental evidence presented in Refs.~\cite{MafiMMI1,MafiMMI2} confirms the reliability of this choice.
Moreover, using the formalism outlined in Ref.~\cite{MafiNL}, it can be shown that the conservation of 
angular momentum, dictated by the symmetries of the nonlinear coupling terms, prevents the excitation of 
any modes with $m\neq 0$ along the fiber from only $m=0$ modes; therefore, 
our analysis { can  be safely limited}
to the small and manageable zero-angular momentum subset of modes. 
{ It is worth mentioning} that $m\neq 0$ modes may still be excited
because of the bending and imperfections in the fiber; this issue will be addressed in Section~\ref{sec:conclusion} but for now, 
our analysis { are limited} to the subspace of $m=0$ modes.

{ Because only ${\rm LG}_{p0}$ modes are considered in the analysis, one} 
can readily identify the collective 
index $\mu$ used in Section~\ref{sec:GeneralOverview} with the radial mode number $p$. 
 
Therefore, all the indexes in the
relevant equations such as in Eqs.~\ref{eq:masterNL},~\ref{eq:gammadef},~\ref{eq:masterNL2}, and~\ref{eq:beamexpansion}
can be directly replaced with the radial mode numbers of the ${\rm LG}_{p0}$ modes. 

In a typical GIMF, only a handful of $m=0$ modes are supported in the core; for example, only five
$m=0$ modes are supported at 1550~nm wavelength in a C-GIMF. Even for such a small number of modes, 
the GNLSE will have many terms and becomes quite complicated. In order to develop the proper
fundamental understanding and intuition on the key mechanisms involved in the NL-MMI of these modes
in a GIMF, 
{ our analysis are initially limited to} the smaller subset of
$\LG_{00}$ and $\LG_{10}$ modes in Subsection~\ref{sec:2modes}. This way, 
{ the physics} 
of the NL-MMI in the SMF-GIMF-SMF is not buried under the complexity introduced by the large number of modes 
propagating in the GIMF. Eventually, in Subsection~\ref{sec:5modes}, 
{ the number of propagating modes in the GIMF will be increased to five, considering $\LG_{00}$ through $\LG_{40}$, and the similarities and differences 
between the two-mode and the five-mode scenarios will be explored.}

As a reference for comparison with nonlinear couplings for 
self-phase modulation and cross-phase modulation in conventional SMFs~\cite{AgrawalNLBook},
{ the nonlinear coupling terms among the three lowest order modes 
in GIMFs; i.e., ${\rm LG}_{00}$, ${\rm LG}_{10}$, and ${\rm LG}_{20}$, are:
$\etat_{0001}=\etat_{0011}=\etat_{1111}=1/2$,
$\etat_{0012}=\etat_{0022}=3/8$,
$\etat_{2222}=11/32$,
$\etat_{1122}=5/16$,
$\etat_{0002}=\etat_{0111}=\etat_{0112}=\etat_{1112}=1/4$,
$\etat_{0122}=3/16$,
$\etat_{1222}=11/64$,
$\etat_{0222}=5/32$.
}
\subsection{NL-MMI with $\LG_{00}$ and $\LG_{10}$ modes}
\label{sec:2modes}
In this subsection, 
{ it is assumed} that only $\LG_{00}$ and $\LG_{10}$ modes are excited in the GIMF and
the optical power is distributed only between these two modes.
{ It should be taken into account that} this assumption is only strictly valid for $\eta\approx 1$  from Eq.~\ref{eq:overlap} {($p_0\gg p_1$ in that case)}.
However, our intention in this section is not to simulate an exact experimentally realizable configuration
in the spirit of Fig.~\ref{fig:coupler};
rather, 
{ this limited two-mode subspace is used to gain insight}
into the physics of NL-MMI in GIMFs. Therefore, in this subsection, 
the limitations of Eq.~\ref{eq:overlap} { are abandoned and it is assumed} that $p_0+p_1=1$, where  $p_0=|B_0|^2$ and $p_1=|B_1|^2$ can adopt any (positive) values subject to this condition. In practice, 
it may be possible to create this scenario experimentally in a more complex setup than that of Fig.~\ref{fig:coupler}; e.g.,
by using mode conversion techniques; however, such details are beyond the scope and
intentions of this subsection.

Using Eq.~\ref{eq:masterNL2}, 
the following coupled nonlinear equations { are obtained}:
\begin{align}
\nonumber
\partial_\zeta B_0&=
i\gammat (|B_0|^2+\dfrac{1}{4}|B_1|^2)B_1
+
\dfrac{i\gammat}{2}(B_0^2B^\star_1+B_1^2B^\star_0),
\\
\nonumber
\partial_\zeta B_1&=-ir_1B_1+\dfrac{i\gammat}{2}B_0\\
&+\dfrac{i\gammat}{2}(-|B_1|^2B_1+B_0^2B^\star_1+\dfrac{1}{2}B_1^2B^\star_0),
\label{eq:masterNL0001}
\end{align}
Nonlinear effects can be ignored when $\gammat \Lt\approx 0$, where 
$\Lt=L\times \left(\beta_{(1)}-\beta_{(2)}\right)$ and $L$ is the total length of the GIMF.
In the absence of nonlinear effects, the solution to Eq.~\ref{eq:masterNL0001} can be written as
\begin{align}
B_0(\zeta)=B_0(0),\qquad 
B_1(\zeta)=e^{-ir_1\zeta}B_1(0).
\label{eq:SolNoNL0001}
\end{align}
{ In the linear case, the relative power of each mode is conserved;} i.e., $p_0$ and $p_1$ remain unchanged along the
GIMF. The relative power transmission in an SMF-GIMF-SMF setup in the linear limit with two modes can be calculated 
from Eq.~\ref{eq:tauSMFGIMFSMF} as
\begin{align}
\tau=1-4p_0p_1\sin^2(\dfrac{r_1 \Lt}{2}).
\label{eq:tauNoNL0001}
\end{align}
The relative power transmission is a periodic function of the length of the GIMF and varies periodically  
between $\tau_{\rm min}=1-4p_0p_1$ and $\tau_{\rm max}=1$; $\tau_{\rm min}=0$ in the special case where the 
power is equally distributed ($p_0=p_1=1/2$). These issues have been discussed in further detail in Ref.~\cite{MafiMMI1}, where the linear MMI in GIMFs 
is explored in the presence of multiple $m=0$ modes.

\begin{figure}[htb]
\centering
\includegraphics[width=2.2in]{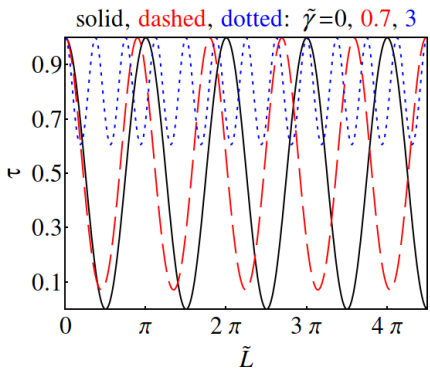}
\caption{The relative power transmission is plotted as a function of the normalized GIMF length $\Lt$
for the case of $\LG_{00}$ and $\LG_{10}$ modes when $p_0=p_1=0.5$ at $\zeta=0$, for $\gammat=0$ (solid), $\gammat=0.7$ (dashed), 
and $\gammat=3$ (dotted).}
\label{fig:tau-2modes-1}
\end{figure}

The main parameters that determine the value of the relative power transmission in the SMF-GIMF-SMF setup are
the relative power of the modes ($p_0$ and $p_1$), the normalized GIMF length ($\Lt$), and the normalized
nonlinear coefficient $\gammat$. In Fig.~\ref{fig:tau-2modes-1}, 
the relative power transmission { is plotted} as a function
of $\Lt$ for three different cases of $\gammat=0$ (solid), $\gammat=0.7$ (dashed), and $\gammat=3$ (dotted);
all cases are plotted for relative power values of $p_0=p_1=0.5$.
The linear case with $\gammat=0$ clearly follows Eq.~\ref{eq:tauNoNL0001} where the oscillation 
periodicity of $\tau$ with respect to $\Lt$ (normalized GIMF length) is
$T_L=2\pi/r_1\approx \pi$. 

When the nonlinear effects are small, the nonlinearity merely results in an additional phase 
mismatch between the $\LG_{00}$ and $\LG_{10}$ modes; e.g., for $\gammat=0.7$, the accumulated nonlinear phase mismatch is 
nearly $\pi$ at $\Lt=4.5\pi$ and is responsible for changing the $\tau=0$ in the linear case to $\tau\approx 1$ for $\gammat=0.7$ in
Fig.~\ref{fig:tau-2modes-1}. 
{ It should be emphasized} that in the presence of nonlinearity, $p_0$ and $p_1$ are not conserved and vary periodically along the GIMF
($\LG_{00}$ and $\LG_{10}$ exchange power); however, unless we explicitly specify their $\zeta$-dependence,
, when refering to $p_0$ and $p_1$, we mean their initial value at the $\zeta=0$ point in the GIMF and refrain from introducing new variables.

The linear and nonlinear cases of $\gammat=0$ and $\gammat=0.7$
differ in another important feature besides the cumulative nonlinear phase mismatch; the minimum value of $\tau$ for 
$\gammat=0.7$ is larger than zero. This 
effect becomes more prominent for larger relative nonlinear coefficients. In fact, as can be seen
for $\gammat=3$ in Fig.~\ref{fig:tau-2modes-1}, the relative power transmission remains above 60\% for all values of $\Lt$. 

\begin{figure}[htb]
\centering
\includegraphics[width=2.2in]{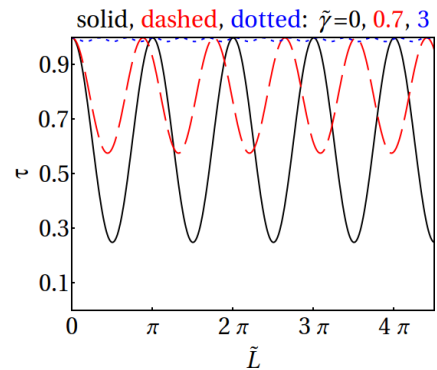}
\caption{Same as Fig.~\ref{fig:tau-2modes-1}, except for $p_0=0.75$ and $p_1=0.25$ at $\zeta=0$.}
\label{fig:tau-2modes-2}
\end{figure}

The relative power transmission explored in Fig.~\ref{fig:tau-2modes-1} is for the case of $p_0=p_1=0.5$.
In Fig.~\ref{fig:tau-2modes-2}, 
similar situations as in Fig.~\ref{fig:tau-2modes-1} { are considered}, yet
with  $p_0=0.75$ and $p_1=0.25$ at $\zeta=0$. The solid line relates to $\gammat=0$ and follows 
Eq.~\ref{eq:tauNoNL0001}. Similar to Fig.~\ref{fig:tau-2modes-1}, the nonlinear phase mismatch between 
the $\LG_{00}$ and $\LG_{10}$ modes increases the frequency of oscillation of $\tau$ with respect of $\Lt$
as the relative nonlinear coefficient is increased. In the case of $\gammat=3$ (dotted), the relative power transmission
$\tau$ remains nearly equal to unity for all values of $\Lt$. For large values of $\gammat$ (e.g. $\gammat=3$),
the power coupling between $\LG_{00}$ and $\LG_{10}$ modes is inefficient and the amplitude of the oscillation 
in the relative power carried by each mode as a function of $\Lt$ is small and is a decreasing function of $\gammat$.
In other words, for a sufficiently large values of $\gammat$, $p_0$ and $p_1$ remain unchanged as they propagate through the GIMF.
The cumulative phase of the $\LG_{00}$ and $\LG_{10}$ modes is also nearly identical for large $\gammat$ as the two modes
propagate along the GIMF. The combined effects of the nearly unchanged relative power and the nearly identical cumulative phase of the modes results in
the near unity value of $\tau$ in this situation.  

\begin{figure}[htb]
\centering
\includegraphics[width=2.2in]{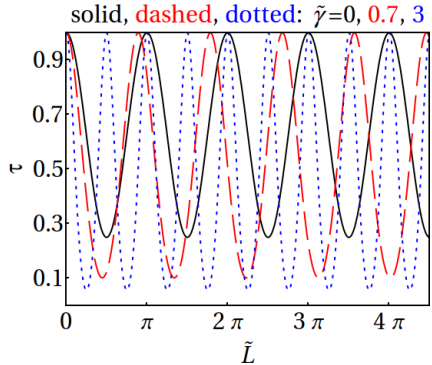}
\caption{Same as Fig.~\ref{fig:tau-2modes-1}, except for $p_0=0.25$ and $p_1=0.75$ at $\zeta=0$.}
\label{fig:tau-2modes-3}
\end{figure}

In Fig.~\ref{fig:tau-2modes-3}, 
similar situations as in Fig.~\ref{fig:tau-2modes-2} { are considered}, yet
with $p_0=0.25$ and $p_1=0.75$ at $\zeta=0$. The solid line relates to $\gammat=0$ and is identical to the
case of $p_0=0.75$ and $p_1=0.25$ presented in Fig.~\ref{fig:tau-2modes-2}, in agreement with Eq.~\ref{eq:tauNoNL0001}.
Similar to Fig.~\ref{fig:tau-2modes-2}, increasing the relative nonlinear coefficient results in a larger
oscillation frequency of $\tau$ with respect to $\Lt$. However, in complete contrast to the behavior observed in 
Fig.~\ref{fig:tau-2modes-2} for $\gammat\neq 0$, increasing $\gammat$ initially lowers the minimum value of $\tau$
and for sufficiently large value of $\gammat$, the minimum can even be equal to zero; if $\gammat$ is further increased,
the minimum value of $\tau$ increases again and eventually settles asymptotically at a value that can be substantially different 
from zero. This behavior is dictated by a very efficient exchange of power between $\LG_{00}$ and $\LG_{10}$ modes as well
as considerable difference between the cumulative phases of the modes; however, at a large $\gammat$ the cumulative phases of 
the modes become nearly identical.

\begin{figure}[htb]
\centering
\includegraphics[width=2.2in]{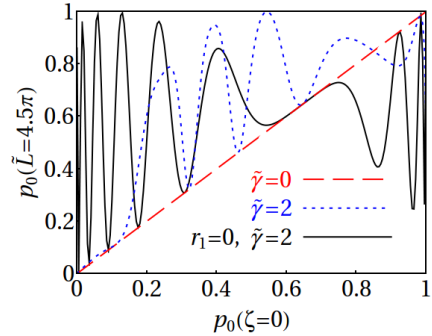}
\caption{This figure shows the exchange of power between the $\LG_{00}$ and $\LG_{10}$ modes, where $p_0$ at the output of 
the GIMF is plotted as a function of $p_0$ at the input for $\Lt=4.5\pi$.}
\label{fig:2modes-p0-p0-45Pi}
\end{figure}

As 
{ discussed above}, the dynamics of power coupling between the $\LG_{00}$ and $\LG_{10}$ modes 
is quite complex. In Fig.~\ref{fig:2modes-p0-p0-45Pi}, 
{ $p_0(\zeta=\Lt)$ is plotted as a function of $p_0(\zeta=0)$ for $\Lt=4.5\pi$.} 
In the linear case, $\LG_{00}$ and $\LG_{10}$ modes are totally uncoupled 
and $p_0(\zeta=\Lt)=p_0(\zeta=0)$, as shown with the diagonal dashed line in Fig.~\ref{fig:2modes-p0-p0-45Pi}. However,
for $\gammat=2$, nonlinearity couples the power between $\LG_{00}$ and $\LG_{10}$ modes and $p_0(\zeta=\Lt)$
oscillates as a function of $p_0(\zeta=0)$, shown with the dotted line in Fig.~\ref{fig:2modes-p0-p0-45Pi}.
{ It can be seen that} $p_0(\zeta=\Lt)$ oscillates more rapidly as $\gammat$ increases. In order to isolate the nonlinear effect,
{ the dispersive term $r_1$ is set to zero and the power coupling curve for $\gammat=2$ is re-plotted}
as the solid line in Fig.~\ref{fig:2modes-p0-p0-45Pi}. Comparing the solid and dotted lines, 
{ it can be concluded that} the effect of the dispersive term ($-ir_1B_1$) is to tame the oscillations, especially for $p_0$ near zero or unity.

\begin{figure}[htb]
\centering
\includegraphics[width=2.2in]{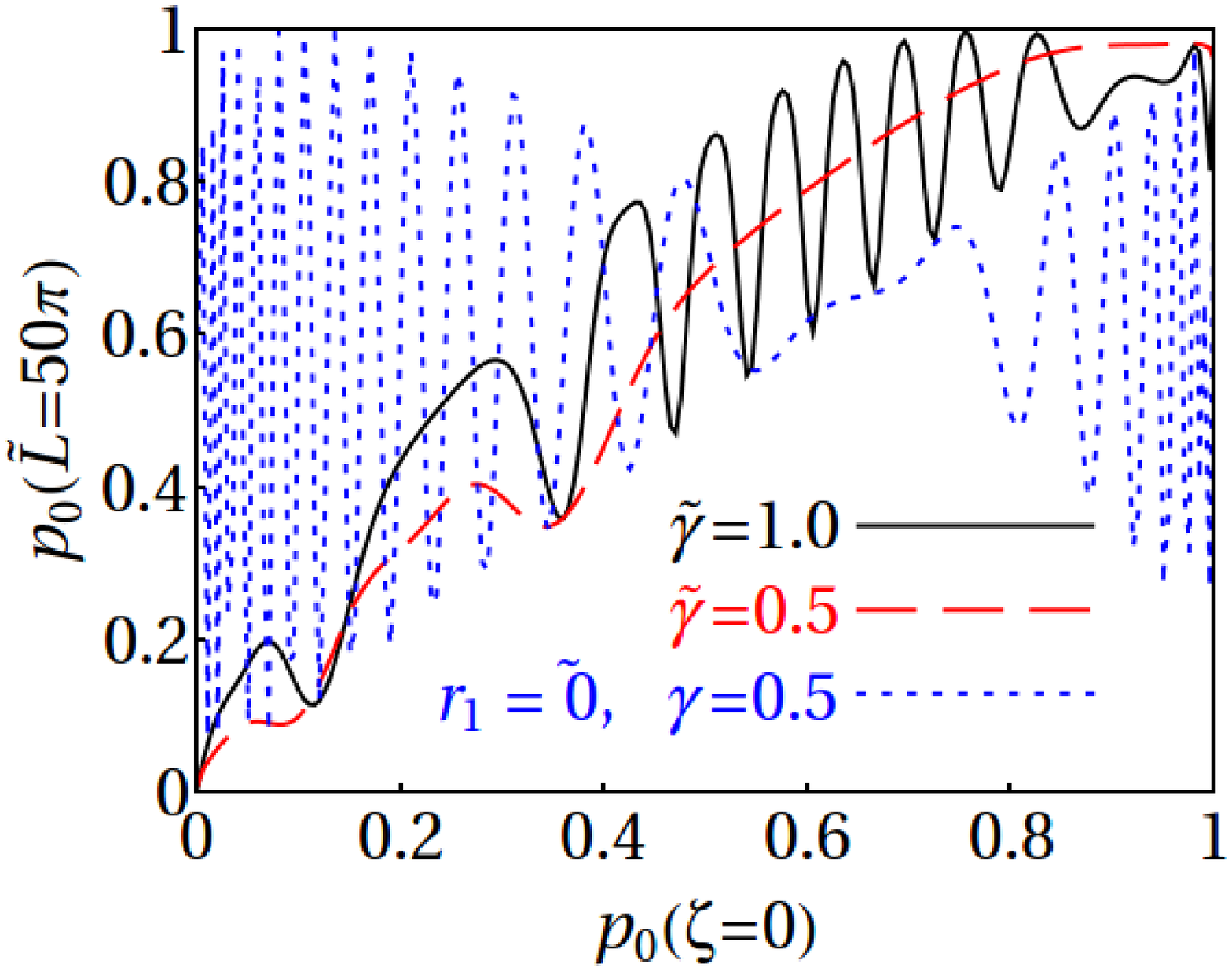}
\caption{Same as Fig.~\ref{fig:2modes-p0-p0-45Pi}, except for $\Lt=50\pi$, and
different choices of $\gammat$.}
\label{fig:2modes-p0-p0-50Pi}
\end{figure}

For a longer GIMF (at fixed $\gammat$), the taming effect of the dispersive term will be even more pronounced and 
the amplitude of oscillations will be smaller. In Fig.~\ref{fig:2modes-p0-p0-50Pi}, 
{ where $\Lt$ is assumed to be $50\pi$, the} ``artificial'' case of $r_1=0, \gammat=0.5$ (dotted) oscillates 
{ rapidly}
and with large amplitude. However, for the real situation where $r_1\approx 2$ (and $\gammat=0.5$), 
the rapid oscillation is replaced with a smooth change (dashed line) that does not move far away from the 
diagonal virtual line of $p_0(\zeta=\Lt)=p_0(\zeta=0)$. Once the nonlinearity is increased to
$\gammat=1.0$, the oscillations reappear, yet with a much lower frequency and smaller amplitude (solid line)
compared with the artificial case of $r_1=0$. It is 
noteworthy that the oscillations in the solid curve are mainly located in the region of the graph
where the dotted curve does not oscillate much.

\begin{figure}[htb]
\centering
\includegraphics[width=2.2in]{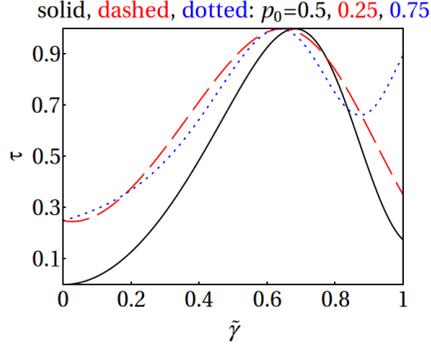}
\caption{
The relative power transmission is plotted as a function of $\gammat$ (thus the total power) for 
$\Lt=4.5\pi$ for the case of $\LG_{00}$ and $\LG_{10}$ modes, when $p_0=0.5, 0.25, 0.75$ at $\zeta=0$,
in solid, dashed, and dotted lines, respectively.}
\label{fig:tau-2modes-4-A}
\end{figure}

In order to examine the effect of the total power injected in the GIMF on the relative power transmission, 
{ the transmission is plotted as a function of} $\gammat$ for the case of $\Lt=4.5\pi$ in Fig.~\ref{fig:tau-2modes-4-A}. The length of the GIMF is chosen in such 
a way that for the linear case of $\gammat=0$, relative power transmission $\tau$ is at its minimum value of 
$\tau_{\rm min}=1-4p_0p_1$ for the three cases of $p_0=0.5, 0.25, 0.75$ shown in Fig.~\ref{fig:tau-2modes-4-A}.
In agreement with our previous discussions, increasing the power (increasing $\gammat$) results in an increase
in the value of $\tau$ and the maximum transmission of $\tau_{\rm max}=1$ is obtained for $\gammat\approx 0.65$,
nearly independent of the initial value of $p_0$. As $\gammat$ is further increased, $\tau$ decreases again and follows an
oscillatory form as a function of $\gammat$. 

\begin{figure}[htb]
\centering
\includegraphics[width=2.2in]{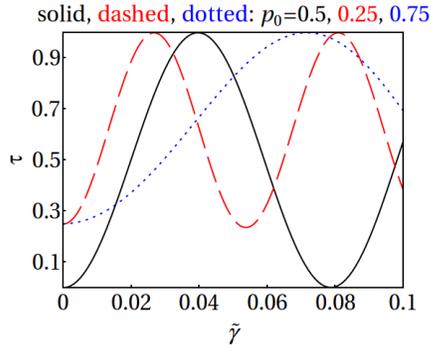}
\caption{Same as Fig.~\ref{fig:tau-2modes-4-A}, except for $\Lt=100.5\pi$.}
\label{fig:tau-2modes-4}
\end{figure}

In practical devices, the GIMF is often considerably longer than $\left(\beta_{(1)}-\beta_{(2)}\right)^{-1}$;
therefore, $\Lt\gg 1$. In Fig.~\ref{fig:tau-2modes-4}, 
a scenario identical to that explored in Fig.~\ref{fig:tau-2modes-4-A} { is considered},
except for a much longer GIMF with $\Lt=100.5\pi$.   
The main difference between the cases of $\Lt=100.5\pi$ in Fig.~\ref{fig:tau-2modes-4} and $\Lt=4.5\pi$ in Fig.~\ref{fig:tau-2modes-4-A} is that
much lower power (much smaller $\gammat$) is required for the GIMF with $\Lt=100.5\pi$
to switch the relative power transmission from $\tau_{\rm min}$ to $\tau_{\rm max}$;
for the case of $p_0=0.25$, the required $\gammat$ for power transmission switching is as low as $\approx 0.026$ as can be seen in the
dashed line in  Fig.~\ref{fig:tau-2modes-4}. Another important difference between the cases of $\Lt=4.5\pi$ and $\Lt=100.5\pi$ is that
in the latter case, the value of $\gammat$ at which 
$\tau_{\rm max}$ is { obtained is} very sensitive to the initial distribution of the power among the modes.

\subsection{Nonlinear MMI with $\LG_{00}$ through $\LG_{40}$ modes}
\label{sec:5modes}
In this subsection, 
the analysis of Subsection~\ref{sec:2modes} { is extended} to the more realistic case
of five propagating modes. As 
{ mentioned before,} this is a realistic scenario for the case of a C-GIMF 
at 1550~nm wavelength. The general observations in this subsection and the similarities and differences
with the two-mode scenario of Subsection~\ref{sec:2modes} should cover the overall nonlinear dynamics in most 
realistic cases of NL-MMI in GIMFs.

In addition to the difference between the number of modes in this subsection versus Subsection~\ref{sec:2modes},
here 
{ the initial excitation amplitudes are chosen such that they satisfy Eq.~\ref{eq:overlap}} (otherwise the problem would become intractable). The main advantage of this choice is that it complies with the
configuration of Fig.~\ref{fig:coupler} and can be tested experimentally.
Another important 
difference is that in Subsection~\ref{sec:2modes}, 
{ it was} assumed that $p_0+p_1=1$, which resulted in $\tau_{\rm max}=1$.
In a realistic case where Eq.~\ref{eq:overlap} is applied and the number of guided modes is finite, some of the power
coming from the input SMF is coupled to the radiation and cladding modes. Therefore, it can be shown (see Ref.~\cite{MafiMMI1}) 
that the maximum transmission is given by
\begin{equation}
\tau_{\rm max}=(1-\Psi^{2P})^2\le 1,
\label{Eq:maxtau}
\end{equation}
where $\Psi$ is defined in Eq.~\ref{eq:overlap} and $P$ is the total number of propagating modes with $m=0$: $P=5$ in this
subsection. Therefore, in all the plots shown in this subsection, $\tau_{\rm max}<1$ and is given by Eq.~\ref{Eq:maxtau}.

\begin{figure}[htb]
\centering
\includegraphics[width=2.2in]{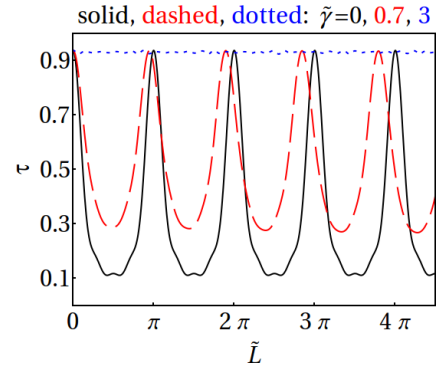}
\caption{The relative power transmission is plotted as a function of the normalized GIMF length $\Lt$
for the case of $\LG_{00}$ through $\LG_{40}$ modes (five zero angular modes) when $p_0=0.5$ at $\zeta=0$ ($\eta=3+\sqrt{8}$), 
for $\gammat=0$ (solid), $\gammat=0.7$ (dashed), and $\gammat=3$ (dotted). The results should be compared with Fig.~\ref{fig:tau-2modes-1}
where only two modes $\LG_{00}$ and $\LG_{10}$ were considered.}
\label{fig:tau-5modes-1}
\end{figure}

In Fig.~\ref{fig:tau-5modes-1}, 
the relative power transmission { is plotted} as a function of the normalized GIMF length $\Lt$,
for $\eta=3+\sqrt{8}$, which results in $p_0=0.5$. Similar to the two-mode scenario plotted in Fig.~\ref{fig:tau-2modes-1},
the cases of $\gammat=0$ (solid), $\gammat=0.7$ (dashed), and $\gammat=3$ (dotted) { are analyzed}. The linear case ($\gammat=0$) in Fig.~\ref{fig:tau-5modes-1}
is very similar to the two-mode case of Fig.~\ref{fig:tau-2modes-1}, except that $\tau_{\rm min}$ and $\tau_{\rm max}$ are different
due to the power leakage to the radiation and cladding modes, as explained above. However, the major difference is that at 
$\Lt=4.5\pi$, $\gammat=0.7$ is not nearly enough to switch the relative power transmission from $\tau_{\rm min}$ (at $\gammat=0$) to 
$\tau_{\rm max}$ in the five-mode scenario. In other words, higher power is required in the realistic five-mode scenario for transmission switching compared with
the two-mode scenario. This is intuitively expected as the presence of the higher modes results in the distribution of the intensity
over a larger cross-sectional area of the GIMF.

\begin{figure}[htb]
\centering
\includegraphics[width=2.2in]{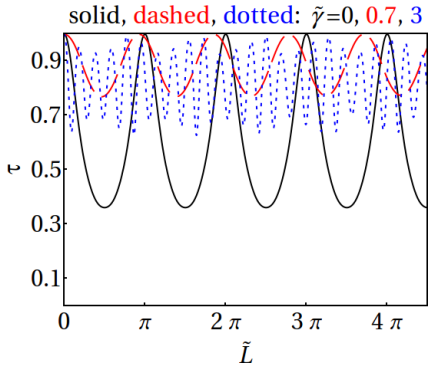}
\caption{Same as Fig.~\ref{fig:tau-5modes-1}, except for $p_0=0.75$ at $\zeta=0$ ($\eta=3$).
The results should be compared with Fig.~\ref{fig:tau-2modes-2}
where only two modes $\LG_{00}$ and $\LG_{10}$ were considered.}
\label{fig:tau-5modes-2}
\end{figure}

{ Next, the case of $p_0=0.75$ for the five-mode scenario is considered in Fig.~\ref{fig:tau-5modes-2},}
which should be compared with the two-mode scenario in Fig.~\ref{fig:tau-2modes-2}. According to Eq.~\ref{eq:overlap},
this case can be obtained by choosing $\eta=3$. Unlike the case of $p_0=0.5$ studied above, there is a strong similarity 
between the five-mode and two-mode scenarios in the $p_0=0.75$ case; at $\Lt=4.5\pi$, $\gammat=0.7$ is nearly sufficient to switch 
the relative power transmission from $\tau_{\rm min}$ (at $\gammat=0$) to $\approx \tau_{\rm max}$. This result is quite important, 
because $\eta=3$ is more readily achievable when using standard commercial fibers than such large values
as $\eta=3+\sqrt{8}\approx 5.8$. For example, 
{ as mentioned before, coupling C-GIMF to SMF-28 at 1550~nm, results in $\eta\approx 2.2$, or coupling 1060XP fiber from Thorlabs catalog to C-GIMF leads to $\eta\approx 4.27$ at 1060~nm wavelength.} The major difference between
the two-mode and five-mode scenarios is that at the higher value of $\gammat=3$, $\tau$ experiences a much higher-amplitude 
oscillation as a function of $\Lt$ in the five-mode case. 
{ The large amplitude oscillations in the five-mode scenario for $\gammat=3$ are due to the efficient power transfer from the 
lowest order mode to the other four modes; each mode accumulates a different linear and nonlinear phase as it propagates
along the GIMF, resulting in large variations in the relative power transmission which is a sensitive function of
both the amplitude and the phase of each mode (see Eq.~\ref{eq:tauSMFGIMFSMF}).}

\begin{figure}[htb]
\centering
\includegraphics[width=2.2in]{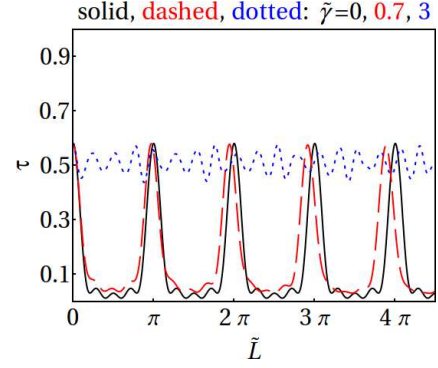}
\caption{Same as Fig.~\ref{fig:tau-5modes-1}, except for $p_0=0.25$ at $\zeta=0$ ($\eta=7+\sqrt{48}$).
The results should be compared with Fig.~\ref{fig:tau-2modes-3}
where only two modes $\LG_{00}$ and $\LG_{10}$ were considered.}
\label{fig:tau-5modes-3}
\end{figure}

Finally, 
the case of $p_0=0.25$ for the five-mode scenario { is considered} in Fig.~\ref{fig:tau-5modes-3}, 
which should be compared with the two-mode scenario in Fig.~\ref{fig:tau-2modes-3}. This case corresponds to 
$\eta=7+\sqrt{48}\approx 13.9$, which is very hard to obtain using conventional fibers. Substantial power
leakage into the radiation and cladding modes is observed in the five-mode scenario and the results contrast sharply from
the two-mode scenario in Fig.~\ref{fig:tau-2modes-3}. 
{ Because of the finite number of core-guided modes supported by the finite GIMF core (see Eq.~\ref{eq:Nalpha}) which 
is truncated by the cladding at radius $R$, some of the power from the input SMF does not couple to the core-guided modes
and couples to the cladding and radiation modes, as estimated in Eq.~\ref{Eq:maxtau} and also discussed in Ref.~\cite{MafiMMI1}.
The power leakage can be substantial for $\eta\gg 1$ as observed in Fig.~\ref{fig:tau-5modes-3}.
}
There is hardly any difference between the linear case ($\gammat=0$) and  $\gammat=0.7$
in Fig.~\ref{fig:tau-5modes-3}; however, for $\gammat=3$, the relative power transmission nearly saturates
over the entire value of $\Lt$ at $\tau_{\rm max}$.  

In summary, the NL-MMI related to $p_0=0.75$ in the five-mode scenario closely resembles that of the two-mode
scenario presented in Subsection~\ref{sec:2modes}. This similarity is lost as the value of $p_0$ is lowered 
(via increasing $\eta$); however, 
{ a larger value} of $p_0$ is generally more accessible using
commercially available optical fibers; fortunately, the power switching dynamics (from $\tau_{\rm min}$ to $\tau_{\rm max}$)
is more desirable for $p_0=0.75$ than $p_0=0.25$ in the SMF-GIMF-SMF geometry shown in Fig.~\ref{fig:coupler}.
\section{Saturable absorber using NL-MMI}
\label{sec:saturable}
In Section~\ref{sec:NLMMI}, 
the NL-MMI behavior of the
SMF-GIMF-SMF geometry of Fig.~\ref{fig:coupler} { was discussed} in great detail. 
Here, 
{ the results will be used} to explore the application of this 
geometry as an SA. Specifically, it is desirable 
for the SMF-GIMF-SMF configuration to attenuate low power signals
($\gammat\approx 0$), but allow the higher power signals to go through.
{ All analysis} performed in this section is for
the realistic case of five-modes ($\LG_{00}$ through $\LG_{40}$),
subject to the constraint of Eq.~\ref{eq:overlap} for the initial excitation amplitudes.

\begin{figure}[htb]
\centering
\includegraphics[width=2.2in]{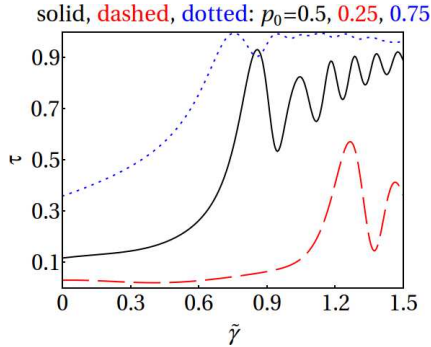}
\caption{
The relative power transmission is plotted as a function of $\gammat$ (thus the total power) for
$\Lt=4.5\pi$ for the case of five modes, when $p_0=0.5, 0.25, 0.75$ at $\zeta=0$,
in solid, dashed, and dotted lines, respectively.}
\label{fig:tau-5modes-4-A}
\end{figure}

{ Fig.~\ref{fig:tau-5modes-4-A} depicts} the behavior of the 
relative power transmission $\tau$ as a function of $\gammat$ for a fixed
value of $\Lt=4.5\pi$ in a SMF-GIMF-SMF configuration, for
$p_0=0.5, 0.25, 0.75$ at $\zeta=0$, in solid, dashed, and dotted lines.
The value of $\Lt$ is chosen such that in the linear case, the
relative power transmission is at its minimum value. In Subsection~\ref{sec:5modes},
{ it was pointed out} that the case of $p_0=0.75$ is the most interesting scenario
from an experimental point of view. For $p_0=0.75$, the relative power 
transmission increases from $\tau_{\rm min}$ to  $\tau_{\rm max}\approx 1$
monotonically as $\gammat$ is increased from zero to $\gammat\approx 0.75$.
Beyond $\gammat\approx 0.75$, $\tau$ goes through a few low amplitude
oscillations and saturates at $\tau_{\rm max}\approx 1$. This is nearly an ideal
scenario for an SA, where the only downside with this design
is that $\tau_{\rm min}$ is considerably larger than zero. 

For $p_0=0.5$ in Fig.~\ref{fig:tau-5modes-4-A}, a slightly higher value of 
$\gammat$ is required compared with $p_0=0.75$ to achieve the maximum
transmission; moreover, more oscillations are observed beyond the transmission
peak value, which may not be a desirable feature for an SA.
Another important feature of this plot is the large low-transmission plateau
and the sudden rise of $\tau$ near the peak values, which is desirable
for pulse shortening in an SA.  
The case of $p_0=0.25$ is also shown in Fig.~\ref{fig:tau-5modes-4-A};
the peak value of transmission in this scenario is only near 50\% due to power
coupling to the radiation and cladding modes, as discussed in Subsection~\ref{sec:5modes}, 
making this an undesirable design.
The results in Fig.~\ref{fig:tau-5modes-4-A} should be compared with the
two-mode scenario in Fig.~\ref{fig:tau-2modes-4-A}. A notable feature
in Fig.~\ref{fig:tau-2modes-4-A} is the large-amplitude oscillations of $\tau$ 
as a function of $\gammat$; this behavior is very different from Fig.~\ref{fig:tau-5modes-4-A}
where $\tau$ was nearly saturated to $\tau_{\rm max}$ beyond $\gammat\approx 0.75$ for $p_0=0.75$. 

\begin{figure}[htb]
\centering
\includegraphics[width=2.2in]{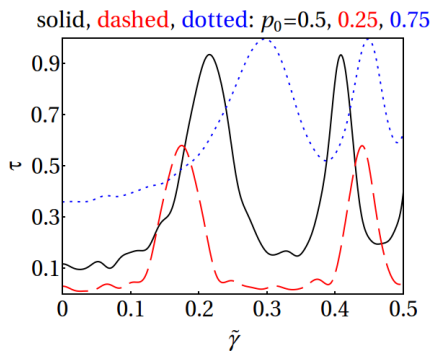}
\caption{Same as Fig.~\ref{fig:tau-5modes-4-A}, except for $\Lt=100.5\pi$.}
\label{fig:tau-5modes-4}
\end{figure}

In Fig.~\ref{fig:tau-5modes-4}, 
{ a similar case to the one discussed in Fig.~\ref{fig:tau-5modes-4-A} is considered,} except for a longer GIMF section
where $\Lt=100.5\pi$. Similarly, this value of $\Lt$ is chosen such that in the linear case, the
relative power transmission is at its minimum value. The most notable differences between
the case of $\Lt=100.5\pi$ in Fig.~\ref{fig:tau-5modes-4} and $\Lt=4.5\pi$ in Fig.~\ref{fig:tau-5modes-4-A} are that
$\tau_{\rm max}$ is obtained at a lower value of $\gammat$ for the longer GIMF design; this is somewhat expected, given
that the nonlinear phase is cumulative. 
{  However, comparing the five-mode scenario in Fig.~\ref{fig:tau-5modes-4} with the two-mode case of Fig.~\ref{fig:tau-2modes-4}, shows that} 
much higher value of $\gammat$ is required in the five-mode scenario to obtain $\tau_{\rm max}$; therefore,
the five-mode and two-mode scenarios differ substantially from each other.

In Figs.~\ref{fig:tau-5modes-4-A} and~\ref{fig:tau-5modes-4}, 
{ the value of $\Lt$ is chosen such that $\tau=\tau_{\rm min}$ is obtained for $\gammat=0$.}
While this may seem a reasonable choice, 
{ however}
this is not required from an SA;
rather, all that is required from an SA is to efficiently discriminate against low
power signals with minimal impact on the high power. Therefore, it is necessary to 
examine other possibilities for $\Lt$ and find the most efficient SA that is possible
with the SMF-GIMF-SMF geometry of Fig.~\ref{fig:coupler}. A relevant metric for the 
pulse power discrimination in an SA is the value of $(\partial\tau/\partial\gammat)$,
which characterizes the sensitivity of an SA to a change in the pulse power.

\begin{figure}[htb]
\centering
\includegraphics[width=2.2in]{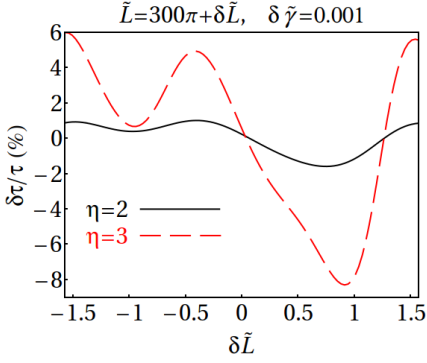}
\caption{
{ The relative change in the} normalized power transmission in the SMF-GIMF-SMF geometry defined in Eq.~\ref{Eq:deltatau} { plotted} as a function of 
$\delta\Lt$, where the length of the GIMF section is given by $\Lt=300\pi+\delta\Lt$. A large positive value of  $(\delta\tau/\tau)$ is desirable 
for saturable absorption. The plots are presented for $\eta=2$ (solid) and $\eta=3$ (dashed).}
\label{fig:5modes-5-A}
\end{figure}

{ Fig.~\ref{fig:5modes-5-A} shows $(\delta\tau/\tau)$, defined as}
\begin{equation}
(\delta\tau/\tau)=\dfrac{1}{\tau}\Big(\tau|_{\gammat=0.001}-\tau|_{\gammat=0}\Big),
\label{Eq:deltatau}
\end{equation}
as a function of the normalized GIMF length. 
{  $(\delta\tau/\tau)$ is used as the metric because it is}
related to $(\partial\tau/\partial\gammat)$ via 
$(\delta\tau/\tau)\approx \tau^{-1}(\partial\tau/\partial\gammat)\delta\gammat$ for $\delta\gammat=0.001$.
The normalized 
length of the GIMF is chosen to be $\Lt=300\pi+\delta\Lt$, and the horizontal axis in Fig.~\ref{fig:5modes-5-A} 
is expressed as $\delta\Lt$ in the range $[-\pi/2,+\pi/2]$; the reason for this limited range in $\delta\Lt$ is that $(\delta\tau/\tau)$ 
repeats in a nearly periodic fashion outside this range, unless $\delta\Lt$ becomes very large. 
{ It will be discussed later in Section~\ref{sec:conclusion} that even} a seemingly small value of $\gammat=0.001$ can translate to multi-kilowatts of peak power in some commercial GIMFs, such as C-GIMF.
Therefore, our main intention in choosing $\delta\gammat=0.001$ in Eq.~\ref{Eq:deltatau}
and subsequent figures is to ensure that our design and observations are relevant for practical situations.

For both $\eta=2$ (solid line) and $\eta=3$ (dashed line) in Fig.~\ref{fig:5modes-5-A}, the maximum value of  $(\delta\tau/\tau)$ is obtained
at $\delta\Lt\approx -0.4$, while the large low-transmission plateau already observed in Figs.~\ref{fig:tau-5modes-4-A} and~\ref{fig:tau-5modes-4}
at near $\gammat=0$ is responsible for the small value of $(\delta\tau/\tau)$ near $\delta\Lt=0$. The maximum value of $(\delta\tau/\tau)$
is around 1\% for $\eta=2$ and 5\% for $\eta=3$. 
{ It should be pointed out that} SAs can be used to mode-lock lasers even when their modulation depth (MD) is as low as
0.5\%~\cite{Kartner}; MD is defined as the maximum change in transmission in an SA. Although MD is often the standard quantity used to asses the performance of SAs,
{ the large values} of MD obtained in the SMF-GIMF-SMF (between $\tau_{\rm min}$ and $\tau_{\rm max}$) can be misleading, because 
{ a very high peak power} may be needed to access the 
full range of the MD 
{ (some numerical values will be presented in Section~\ref{sec:conclusion}).}
Therefore, 
{ $(\delta\tau/\tau)$ has been used in this paper,} 
which is identical to MD within the reasonably accessible range of pulse peak powers.
 
\begin{figure}[htb]
\centering
\includegraphics[width=2.2in]{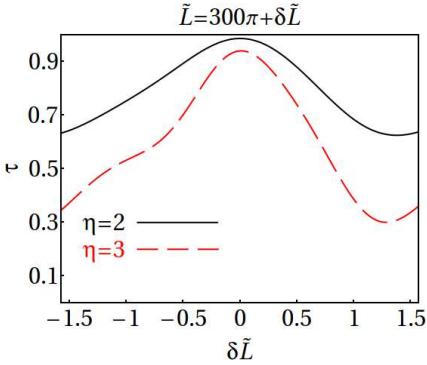}
\caption{The relative power transmission is shown as function of $\delta\Lt$, where the length of the GIMF section is given by $\Lt=300\pi+\delta\Lt$.
$\tau$ presented in this figure is the average value between $\tau|_{\gammat=0.001}$ and $\tau|_{\gammat=0}$. This figure is intended to show that
the value of $\tau$ is not too low near the peak value of $(\delta\tau/\tau)$ in Fig.~\ref{fig:5modes-5-A}. The plots are presented for $\eta=2$ (solid) 
and $\eta=3$ (dashed).}
\label{fig:5modes-5-B}
\end{figure}

{ It is required to} ensure that the total transmission through the SMF-GIMF-SMF geometry is not too low; in other words,
the SA is not too lossy near the peak value of $(\delta\tau/\tau)$. In Fig.~\ref{fig:5modes-5-B}, 
the average value $(\tau|_{\gammat=0.001}+\tau|_{\gammat=0})/2$ { is plotted} as a function of $\delta\Lt$ for $\eta=2$ (solid line) 
and $\eta=3$ (dashed line). It can be seen that the average relative power transmission is reasonably large near the peak value of 
$(\delta\tau/\tau)$ of Fig.~\ref{fig:5modes-5-A}; therefore, the SA can be regarded as a viable design.

\begin{figure}[htb]
\centering
\includegraphics[width=2.2in]{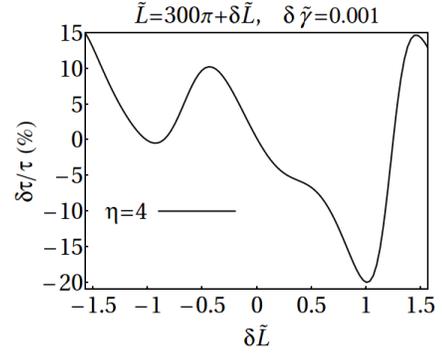}
\caption{Same as Fig.~\ref{fig:5modes-5-A}, except for $\eta=4$.}
\label{fig:5modes-5-AA}
\end{figure}

\begin{figure}[htb]
\centering
\includegraphics[width=2.2in]{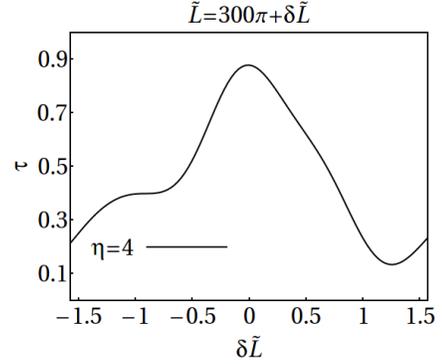}
\caption{Same as Fig.~\ref{fig:5modes-5-B}, except for $\eta=4$.}
\label{fig:5modes-5-BB}
\end{figure}

In Fig.~\ref{fig:5modes-5-AA}, 
{ $(\delta\tau/\tau)$ is plotted} in the same fashion as in Fig.~\ref{fig:5modes-5-A}, except for $\eta=4$. 
{ The reason separate figures are used}
to plot $\eta=2, 3$ and $\eta=4$ is that the vertical scales are different. As can be seen in Fig.~\ref{fig:5modes-5-AA}, 
$(\delta\tau/\tau)$ as large as 15\% is possible in this case. While this may seem like a highly desirable SA, the relative power transmission
plot for $\eta=4$ in Fig.~\ref{fig:5modes-5-BB} shows that $\tau$ is quite small near the peak values of $(\delta\tau/\tau)$. Perhaps an optimum 
operation of this device can be accomplished at $\delta\Lt\approx-0.3$ where $\delta\tau/\tau\approx 9\%$.
\section{Discussion and Conclusion}
\label{sec:conclusion}
In this paper, 
{ a detailed analysis of the NL-MMI behavior of the SMF-GIMF-SMF geometry shown in 
Fig.~\ref{fig:coupler} has been presented}. Recently, this setup was successfully used to create very low-loss couplers between two SMFs 
with very different mode-field diameters~\cite{MafiMMI1,MafiMMI2}. 
Here, 
{ it has been shown} that the transmission through this coupler has a very interesting nonlinear behavior
and can be potentially used for switching purposes or for saturable absorption in all-fiber mode-locked lasers. 

The key parameters that affect the NL-MMI in this geometry are the ratio of the mode-field diameter of the
$\LG_{00}$ mode in GIMF to the mode-field diameters of the SMFs (characterized by $\eta$ defined in Eq.~\ref{eq:overlap});
the length of the GIMF section (characterized by the relative length  parameter $\Lt$); the relative nonlinear 
coefficient $\gammat$, which incorporates the total power and nonlinear coefficient of the GIMF; and the total
number of propagating modes in the GIMF. 

{ It was observed} that the best performances for switching and intensity discrimination applications such as in SAs can be obtained 
when the value of $\eta$ defined in Eq.~\ref{eq:overlap} is not too large; $\eta<4$ seems like a reasonable choice.
This is easily achievable for reasonable selection of commercially available fibers.

Like most other nonlinear devices, the interaction length plays a very important role in the behavior of the SMF-GIMF-SMF 
device. 
{ The dimensionless normalized length $\Lt$ has been used} throughout this paper, which is the length of the GIMF normalized by 
$\beta_{(1)}-\beta_{(2)}$. This normalization factor is $(183~\mu m)^{-1}$ for a C-GIMF operating at 1550 nm. Therefore,
$\Lt=4.5\pi$, $\Lt=100.5\pi$, and $\Lt=300\pi$ used in this paper translate to $L$=2.6~mm, 5.78~cm, and 17.24~cm
for the length of the GIMF section, which are all reasonable lengths for device applications. For SA purposes,
{ $\Lt=300\pi+\delta\Lt$ was explored} in Figs.~\ref{fig:5modes-5-A} and~\ref{fig:5modes-5-AA}, where it 
{ was realized that}
some level of fine-tuning of the order of $\delta\Lt\sim \pi$ is required, which is  $\delta L\sim$575~$\mu m$
for C-GIMF at 1550~nm wavelength. This situation is not different from that of Refs.~\cite{MafiMMI1,MafiMMI2}, and the fact 
that the required fine-tuning range is so small means that it can be easily achieved by polishing the GIMF to the desired 
length. 
{ It should be emphasized} that the problem of fine-tuning is not very serious and the length needs to be adjusted over a maximum one period
of $\delta\Lt\sim \pi$, because reducing $\Lt$ to $\Lt-\pi$ shifts the periodic transmission pattern by one full period.
Other pragmatic approaches can be taken instead of fine-tuning the length. For example, an ensemble of twenty 
SMF-GIMF-SMF couplers can be built around the same average GIMF length; using this approach, the value of 
$\delta\Lt$ will be uniformly distributed over the range of $[-\pi/2,+\pi/2]$ and one of the elements of the ensemble
will likely perform up to the required specification. 

The relative nonlinear coefficient $\gammat$ is defined in Eq.~\ref{eq:rmugammat}. For C-GIMF at 1550~nm wavelength,
$\gammat$=1.0 corresponds to a power of $\Pt\approx$10.0~MW, given that the nonlinear coefficient $\gamma$ 
(defined for the $\LG_{00}$ mode in GIMF) is smaller than that of the SMF-28 by a factor $\eta\approx 2.2$. For practical 
fiber-based mode-locking applications, this value is unacceptably large. The feasible range of operation for the peak-power 
of a mode-locked fiber laser is in the several to a likely maximum of a few hundred KW of peak power. If the GIMF is made
from highly nonlinear material, e.g., chalcogenide glass with a nonlinear coefficient of up to 1000 times larger than
silica, then $\gammat$=1.0 will correspond to $\Pt\approx$10.0~KW

{ The feasibility of using the SMF-GIMF-SMF geometry has been shown for SA applications even for}
$\gammat=0.001$ in Fig.~\ref{fig:5modes-5-AA}, which corresponds to $\Pt\approx$10.0~KW using C-GIMF and SMF-28 commercial 
fibers at 1550~nm wavelength. Therefore, 
{ it is probable}
that the SMF-GIMF-SMF geometry can be used even with conventional 
commercially available fibers to mode-lock fiber lasers at the presently achievable power levels. The main advantage of 
this geometry is that it can be designed to operate at a much higher power level compared with the existing SAs that exploit 
nonlinear polarization rotation, semiconductors, or carbon nanotubes. Therefore, it might offer a very attractive solution 
for scaling up the pulse energy and peak power in mode-locked fiber lasers that are currently limited by the existing SA 
technology.
 
Throughout this paper, 
{ it was observed} 
that having a smaller number of propagating modes in the GIMF is beneficial, especially in the design of 
an SA. While specific designs need to be carefully analyzed using the formalism laid out in this manuscript, this can be
taken as a general guideline and highly multimode GIMFs should be avoided if possible. We do not think 
that the quality and the telecommunication bandwidth of the GIMF (e.g., due to the centerline defect~\cite{Mafibandwdith}) 
has any appreciable effect on the NL-MMI behavior and the SA performance of the SMF-GIMF-SMF geometry.

Finally, we would like to comment on the possibility of exciting nonzero angular momentum modes, which are excluded from our analysis. 
{ It is possible} to excite $m=\pm 1$ modes via coupling them to $m=0$ modes by bending the GIMF. In order for the
power exchange to operate efficiently in a resonant fashion, the phase mismatch between the $m=0$ and $m=\pm 1$ must be compensated for
by the bend; e.g, long-period gratings are optimized to provide this phase-matching condition~\cite{Ramachandran1}. However, in a GIMF, 
the $m=0$ and $m=\pm 1$ modes always belong to different mode groups with a very large mismatch in their propagation constants
according to Eq.~\ref{eq:beta0th}; 
therefore, a very small bending radius (submillimeter for C-GIMF) is required
for efficient phase matching. Therefore, coupling between $m=0$ and $m=\pm 1$ will not be an efficient process in GIMFs, unless 
assisted by specialty devices such as long-period gratings~\cite{Ramachandran1}. 

Coupling between between $m=0$ and $m=\pm 2$
is induced via the random variations in the diameter of the GIMF, caused by manufacturing errors~\cite{Olshansky1}, and can even happen within the same mode group
with minimal phase matching issue. While this can be minimized by selecting a high quality GIMF, in practice, this should not cause any 
issues for GIMFs shorter than approximately one meter in length. 

{ It has also been previously shown that fabrication errors due to the misalignment of the fibers can be minimized to the extent where
they do not have a noticeable impact on the linear operation of the device~\cite{MafiMMI2}; similar robustness is expected in the nonlinear regime of operation.} 
\section*{Acknowledgments}
The authors acknowledge support from the Air Force Office of Scientific Research under Grant FA9550-12-1-0329.



\end{document}